\documentclass[MSNbibl,number,citesort,seceqn,dvips]{arxbj}
\usepackage{upgreek}
\usepackage{graphicx}

\aid{0}
\volume{18}
\issue{4}
\pubyear{2012}
\firstpage{1223}
\lastpage{1248}
\doi{10.3150/11-BEJ383}

\makeatletter
\newtheorem{prop}{Proposition}[section]

\makeatother

\begin{document}
\begin{frontmatter}

\title{$\varepsilon$-Strong simulation of the Brownian path}
\runtitle{$\varepsilon$-Strong simulation of the Brownian path}

\begin{aug}
\author[a]{\fnms{Alexandros} \snm{Beskos}\corref{}\thanksref{a}\ead[label=e1]{alex@stats.ucl.ac.uk}},
\author[b]{\fnms{Stefano} \snm{Peluchetti}\thanksref{b}\ead[label=e2]{stefano.peluchetti@hsbcib.com}}
\and
\author[c]{\fnms{Gareth} \snm{Roberts}\thanksref{c}\ead[label=e3]{gareth.o.roberts@warwick.ac.uk}}
\runauthor{A. Beskos, S. Peluchetti and G. Roberts} 
\address[a]{Department of Statistical Science, UCL, Gower Street,
London, WC1E 6BT,
UK.\\
\printead{e1}}

\address[b]{HSBC Bank, 8 Canada Square, London, E14 5HQ, UK.
\printead{e2}}

\address[c]{Department of Statistics, University of Warwick, Coventry,
CV4 7AL,
UK.\\
\printead{e3}}
\end{aug}

\received{\smonth{11} \syear{2009}}
\revised{\smonth{1} \syear{2011}}

%
\begin{abstract}
We present an iterative sampling method which delivers upper and lower
bounding processes for the Brownian path. We develop such processes
with particular emphasis
on being able to unbiasedly simulate them on a personal computer.
The dominating processes converge almost surely in the supremum and
$L_1$ norms.
In particular, the rate of converge in $L_1$ is of the order $\mathcal
{O}(\mathcal{K}^{-1/2})$,
$\mathcal{K}$ denoting the computing cost. The a.s. enfolding of the
Brownian path can be exploited in
Monte Carlo applications involving Brownian paths
whence our algorithm (termed the $\varepsilon$-strong algorithm) can
deliver \textit{unbiased} Monte Carlo
estimators over path expectations, overcoming discretisation errors
characterising standard approaches.
We will show analytical results from applications of the $\varepsilon
$-strong algorithm for estimating
expectations arising in option pricing.
We will also illustrate that individual steps of the algorithm can be
of separate interest,
giving new simulation methods for interesting Brownian distributions.
\end{abstract}

%

\end{frontmatter}

\section{Introduction}
Brownian motion (BM) is an object of paramount significance in
stochastic modelling.
Starting from its original mathematical formulation by~\cite{bach00},
its properties are still
under meticulous investigation by contemporary researchers.
Relevant to the purposes of this paper, considerable work has focused on
various constructions and representations
of BM paths. Leaving aside the simple finite-dimensional Gaussian structure
of BM, researchers have often been interested on more complex functionals.
Hitting times, extremes, local times, reflections and other
characteristics of
BM have been investigated (for a general exposition see~\cite{revu99}).
For simulation purposes, many of the relevant distributions are easy to
sample from on a computer~\cite{devr86}.
Several \textit{conditioned} constructions of BM are also known relating
BM with the Bessel process,
the Rayleigh distribution and other stochastic objects (see, e.g.,
\cite{bert99}).

This paper presents a contribution of our own at simulation methods for
Brownian dynamics.
We develop an iterative sampling algorithm, the $\varepsilon$-strong
algorithm, which
simulates upper and lower paths enveloping a.s. the Brownian path.
To meet this objective, we collect a number of characterisations and
combine them in a way
that they can deliver simple sampling methods implementable on a
personal computer.
We will show that after $\mathcal{O}(\mathcal{K})$-computational
effort, the dominating
process have $L_1$-distance of $\mathcal{O}(\mathcal{K}^{-1/2})$.
This a.s. enfolding of the Brownian path can be exploited in
Monte Carlo applications involving Brownian motion integrals, minima,
maxima or hitting times;
in such scenaria, the $\varepsilon$-strong algorithm can deliver unbiased
Monte Carlo
estimators over Brownian expectations, overcoming discretization errors
characterising standard approaches (for the latter approaches, see, for
instance, the exposition in~\cite{glas04}
in the context of applications in finance).

We will show applications of the algorithm and experimentally compare
the required computing resources
against typical alternatives employed in the literature involving Euler
approximation.
Our examples will involve a collection of double-barrier option pricing
problems in a Black and Scholes
framework arising in finance.
Also, we will demonstrate that individual steps of the algorithm can be
of separate interest,
giving new simulation methods for interesting Brownian distributions.

The $\varepsilon$-strong algorithm delivers a pair of dominating
processes, denoted by
$X^{\downarrow}(n)=\{ X^{\downarrow}_{u}(n);u\in[0,1]\}$ and
$X^{\uparrow}(n)=\{X^{\uparrow}_{u}(n);u\in
[0,1]\}$, that can be simulated on a personal computer without
any discretisation error, with the property:
%
%
\begin{equation}
\label{eqdominate}
X^{\downarrow}_{u}(n) \le X^{\downarrow}_{u}(n+1) \le X_u \le
X^{\uparrow}_{u}(n+1) \le X^{\uparrow}_{u}(n)
\end{equation}
for all instances $u\in[0,1]$; here, $X$ is the Brownian path.
The two dominating processes will converge in the limit:
%
%
\begin{equation}
\label{eqconverge}
\mathrm{w.p.1} ,\qquad\lim_{n\rightarrow\infty}
\sup_{u\in[0,1]}\vert X^{\uparrow}_{u}(n)-X^{\downarrow
}_{u}(n)\vert\rightarrow0 .
\end{equation}
The algorithm builds on the notion of the \textit{intersection layer},
a collective information, containing the starting and ending points of
a Brownian path
together with information about its extrema.
A~number of operations (bisection, refinement, see main text) can be
applied on this information,
explicitly on a computer,
allowing the sampler to iterate itself to get closer to $X$.

We should note here that the methods described in this paper will be relevant
also for nonlinear Stochastic Differential Equations (SDEs).
Recent developments in the simulation of SDEs under the framework of
the so-called `Exact Algorithm'
(see~\cite{besk05b,besk06,besk06b,jour07,besk08,case08})
build upon
the result that, conditionally on a collection of randomly sampled points,
the path of the SDE is made of independent Brownian paths.
Once this collection of points is sampled, the methodology of this paper
can then be applied separately on each of the constituent Brownian sub-paths.

The structure of the paper is as following. In Section~\ref{seclayer},
we present the notion of the \textit{intersection layer}
which will be critical for our methods. In Section~\ref{secstrong}, we
present the individual steps forming
the $\varepsilon$-strong algorithm; they will require original simulation
techniques for some Brownian distributions.
Once we identify in Section~\ref{secz} the \textit{$\zeta
$-function}, an
alternating monotone series at the core of Brownian dynamics, we
exploit its structure in Section~\ref{secdetails} to analytically
develop these new sampling methods.
In Section~\ref{secexpaths}, we apply the $\varepsilon$-strong algorithm
to unbiasedly estimate some path
expectations arising when pricing options in finance.
We will contrast the computational cost of the algorithm with Euler
approximation alternatives
to get a better understanding of its practical competitiveness.
In Section~\ref{secexother}, we sketch some other potential
applications of the $\varepsilon$-strong algorithm.
We finish with some discussion and conclusions in Section~\ref{secconclusion}.

\section{Intersection layer and operations}
\label{seclayer}
We will, in general, write paths as $X=\{X_u;u\in[s,t]\}$ for $s<t$.
A Brownian bridge on $[s,t]$ is a Brownian motion conditioned to start at
$X_s$ and end at $X_t$, for some prespecified $X_s$, $X_t$;
its finite-dimensional dynamics are easily derivable following this
interpretation (see, for instance,~\cite{revu99}).

\begin{figure}[t!]

\includegraphics{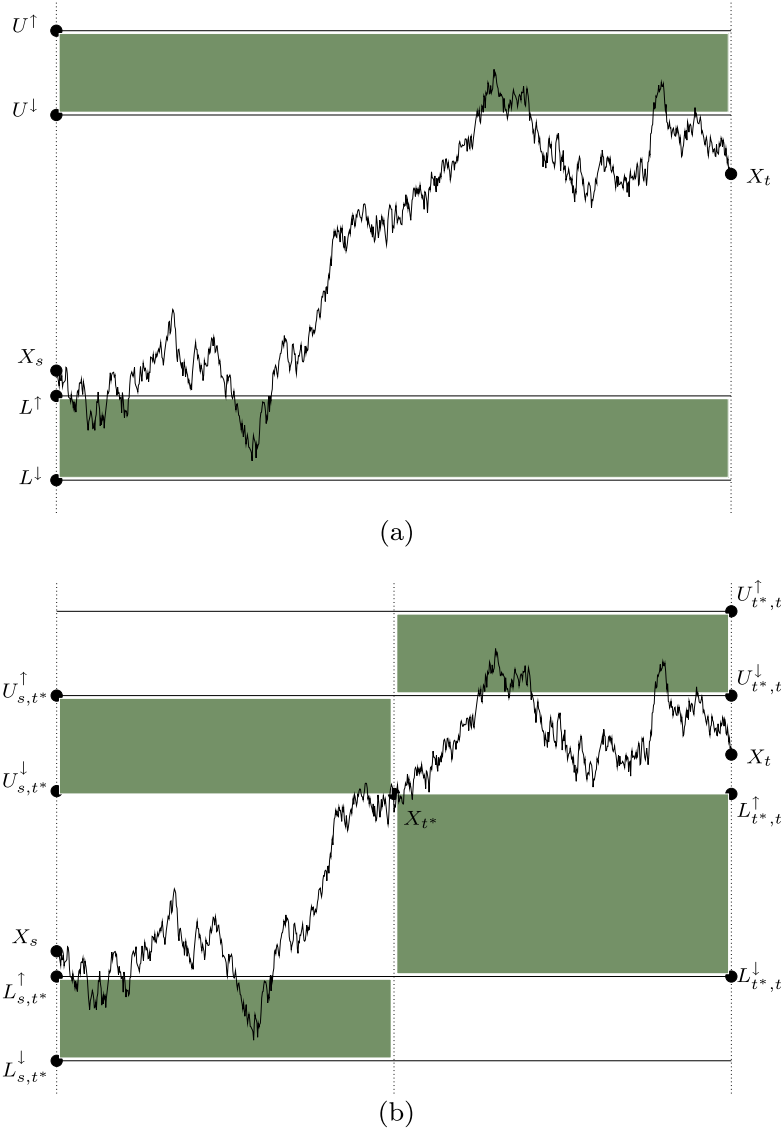}

  \caption{Top panel (a): the intersection layer information $\mathcal{I}_{s,t}$
for a Brownian path. The underlying trajectory starts at $X_s$ and
finishes at $X_t$
with its extrema found in the shaded areas.
Bottom panel (b):
the bisection of $\mathcal{I}_{s,t}$ into $\mathcal{I}_{s,t^*}$ and
$\mathcal{I}_{t^*,t}$.
The algorithm simulates $X_{t^{*}}$ and then decides that
the extrema for each of the intervals $[s,t^{*}]$ and $[t^{*},t]$ are
in the shaded areas, that is, $\mathcal
{U}_{s,t^{*}}=[X_{t^{*}},U^{\downarrow}]$, $\mathcal
{U}_{t^*,t}=[U^{\downarrow},U^{\uparrow}]$, $\mathcal
{L}_{s,t^*}=[L^{\downarrow},L^{\uparrow}]$
and $\mathcal{L}_{t^*,t}=[L^{\uparrow},X_{t^*}]$. The algorithm
outputs the upgraded
information
$\mathcal{I}_{s,t^{*}}=\{X_s,X_{t^*},\mathcal{L}_{s,t^*},\mathcal
{U}_{s,t^*}\}$ and
$\mathcal{I}_{t^*,t}=\{X_{t^*},X_t,\mathcal{L}_{t^*,t},\mathcal
{U}_{t^*,t}\}$.}\label{figbisect}
\end{figure}

Instrumental in our considerations is the notion of (what we call) the
\textit{intersection layer}.
Consider a Brownian bridge $X$ on $[s,t]$. Let $m_{s,t}$, $M_{s,t}$ be
the extrema of $X$:
\[
m_{s,t} = \inf\{X_u;u\in[s,t]\} ,\qquad M_{s,t} = \sup\{X_u;u\in[s,t]\}
.
\]
The $\varepsilon$-strong algorithm will require some information on both
$m_{s,t}$ and $M_{s,t}$.
We will identify intervals:
\[
\mathcal{U}_{s,t}=[U^{\downarrow}_{s,t},U^{\uparrow}_{s,t}] ,\qquad
\mathcal{L}_{s,t}=[L^{\downarrow}_{s,t},L^{\uparrow}_{s,t}] ,
\]
such that:
\[
M_{s,t}\in\mathcal{U}_{s,t} ,\qquad m_{s,t}\in\mathcal{L}_{s,t} .
\]
We will write simply $m$, $M$, $U^{\uparrow}$, $U^{\downarrow
}$, $L^{\uparrow}$, $L^{\downarrow}$
ignoring the
$s,t$-subscripted versions when the time interval under consideration
is clearly implied by the context.
The intersection layer idea refers to the collective information
%
%
\begin{equation}
\mathcal{I}_{s,t}=\{X_s,X_t,\mathcal{L}_{s,t},\mathcal{U}_{s,t}\} ,
\end{equation}
that is the starting and ending points of the bridge together with
intervals that contain its maximum and minimum.
Figure~\ref{figbisect}(a) presents a graphical illustration of the
intersection layer:
the extrema of an underlying Brownian bridge lie in the shaded rectangles.
We will look now at two simple operations on the information $\mathcal
{I}_{s,t}$ which nonetheless
will be the building blocks of the complete $\varepsilon$-strong algorithm
described in the next section.

%
%
\subsection{Refining the information $\mathcal{I}_{s,t}$}
\label{secrefine}
%
%

During the iterations at the execution of
the $\varepsilon$-strong algorithm, for each piece of information
$\mathcal{I}_{s,t}$
we will need to control the width of the layers
$\mathcal{L}_{s,t}$, $\mathcal{U}_{s,t}$ relatively to the size $t-s$
of the time interval
to ensure convergence of the bounding paths enveloping the underlying
Brownian path.
Thus, the refinement of the information $\mathcal{I}_{s,t}$
corresponds to a
procedure that updates $\mathcal{I}_{s,t}$
by halving the allowed width for the minimum $m$ or the maximum $M$ of
the path,
thereby correspondingly updating the layers $\mathcal{L}_{s,t}$ or
$\mathcal{U}_{s,t}$.

More analytically, refinement of $\mathcal{I}_{s,t}$ corresponds to deciding
whether the minimum $m$ on $[s,t]$,
already known to be in $[L^{\downarrow},L^{\uparrow}]$, lies in
$[L^{\downarrow},(L^{\downarrow}+L^{\uparrow})/2]$ or
$[(L^{\downarrow}+L^{\uparrow})/2,L^{\uparrow}]$,
that is, whether $\mathcal{L}_{s,t}$ is equal to $[L^{\downarrow
},(L^{\downarrow}+L^{\uparrow})/2]$
or $[(L^{\downarrow}+L^{\uparrow})/2,L^{\uparrow}]$;
the apparent analogue of such a consideration applies for the maximum $M$.
The analytical method of sampling the relevant binary random variables for
carrying out this procedure will be described in Section~\ref{secdetails}.

%
\subsection{Bisecting the information $\mathcal{I}_{s,t}$}
\label{secbisect}

This is a more involved operation on $\mathcal{I}_{s,t}$, and involves
bisecting
$\mathcal{I}_{s,t}$ into the more analytical information $\mathcal
{I}_{s,t^{*}}\vee
\mathcal{I}_{t^{*},t}$
for some intermediate time instance $t^{*}\in(t,s)$. In particular, we
will be selecting
$t^{*}=(t+s)/2$ within the $\varepsilon$-strong algorithm.
The method begins by sampling the middle point $X_{t^{*}}$ conditionally
on $\mathcal{I}_{s,t}$, and then appropriately sampling the layers for the
two pieces of information
$\mathcal{I}_{s,t^{*}}$, $\mathcal{I}_{t^{*},t}$. The practicalities of
implementing the second part
of the method will depend on whether $X_{t^*}$ falls within a layer of
$\mathcal{I}_{s,t}$ or not,
thus we present the bisection operation in more detail in Table \ref
{tabbisect}.

%
\begin{table}
\caption{The procedure for bisecting the information $\mathcal
{I}_{s,t}$. It
returns the intersection layers
$\mathcal{I}_{s,t^*}$ and $\mathcal{I}_{t^*,t}$ with refined
information about the
underlying path (compared to $\mathcal{I}_{s,t}$)}
\label{tabbisect}
\begin{tabular*}{\textwidth}{@{\extracolsep{\fill}}l@{}}
\hline
\textit{Bisect($\mathcal{I}_{s,t}$):}\\
\phantom{a}\quad{1.}
Set $t^{*}=(t+s)/2$. Simulate $X_{t^{*}}$ given $\mathcal{I}_{s,t}$.
Set $U^{\downarrow} = U^{\downarrow}\vee X_{t^*}$, $L^{\uparrow
} = L^{\uparrow}\wedge X_{t^*}$.\\
\quad{2a.}
Decide if
$\mathcal{U}_{s,t^{*}} =[X_{s}\vee X_{t^*},U^{\downarrow}]$ or
$[U^{\downarrow},U^{\uparrow}]$.\\
\quad{2b.}
Decide if
$\mathcal{U}_{t^*,t} =[X_{t^*}\vee X_{t},U^{\downarrow}]$ or
$[U^{\downarrow},U^{\uparrow}]$.\\
\quad{2c.}
Decide if
$\mathcal{L}_{s,t^{*}}=[L^{\downarrow}, L^{\uparrow}]$ or
$[L^{\uparrow}, X_{s}\wedge X_{t^*}]$.\\
\quad{2d.}
Decide if
$\mathcal{L}_{t^*,t}=[L^{\downarrow}, L^{\uparrow}]$ or
$[L^{\uparrow}, X_{t^*}\wedge X_{t}]$.
\\
\phantom{a}\quad{3.} Return $\mathcal{I}_{s,t^{*}}\vee\mathcal
{I}_{t^*,t}$.\\
\hline
\end{tabular*}
\end{table}

Note that if $X_{t^*}>U^{\downarrow}$ the two upper layers (for
$\mathcal{I}_{s,t^*}$
and $\mathcal{I}_{t^*,t}$) will be directly
set to $[X_{t^{*}},U^{\uparrow}]$, and we will have to simulate extra
randomness about the underlying path only to determine the lower
layers. Correspondingly, if $X_{t^*}<L^{\uparrow}$ the two lower
layers will
immediately
be set to $[L^{\downarrow},X_{t^{*}}]$. In the scenario when
$L^{\downarrow}<X_{t^*}<U^{\downarrow}$, we will have to
simulate extra randomness to determine all four layers.
We describe in Section~\ref{secdetails}
the algorithms for sampling $X_{t^*}$ and determining the layers.
Figure~\ref{figbisect} shows a graphical illustration of the bisection
procedure.
%
\section{\texorpdfstring{$\varepsilon$-Strong simulation of Brownian path}{epsilon-Strong simulation of Brownian path}}
\label{secstrong}

We introduce an iterative simulation algorithm with input
a Brownian bridge $X$ on the domain $[0,1]$ and output, after $n$ iterations,
upper and lower dominating processes $X^{\downarrow}(n)=\{
X^{\downarrow}_{u}(n);u\in[0,1]\}$ and\vadjust{\goodbreak}
$X^{\uparrow}(n)=\{X^{\uparrow}_{u}(n);u\in[0,1]\}$ satisfying
the monotonicity and
limiting requirements~(\ref{eqdominate})
and (\ref{eqconverge}) respectively.
Note that $X$ here is a continuous time
Brownian bridge path, thus an infinite-dimensional random variable.
However, the bounding processes will be piece-wise constant, thus
inherently finite-dimensional. One will be able to realise
complete sample paths of $X^{\downarrow}(n)$
or $X^{\uparrow}(n)$ on a computer without retreating to any sort of
discretization or approximation
errors (apart from those due to finite computing accuracy).

\subsection{\texorpdfstring{$\varepsilon$-Strong algorithm}{epsilon-Strong algorithm}}
\label{secalg}
Given some initial intersection layer information $\mathcal{I}_{0,1}$,
the algorithm will naturally set $X^{\uparrow}_{u}(0)=U^{\uparrow}_{0,1}$
and $X^{\downarrow}_{u}(0)=L^{\downarrow}_{0,1}$ for all instances
$u\in[0,1]$. It will then
iteratively \textit{bisect} the acquired intersection layers, as
described in Section~\ref{secbisect}, to
obtain more information about the underlying sample path on
finer time intervals.
To ensure convergence of the discrepancy
$X^{\uparrow}(n)-X^{\downarrow}(n)$ the algorithm will
sometimes \textit{refine} the information
on some intersection layers, as described in Section~\ref{secrefine},
to reduce the uncertainty for the extrema. We give the pseudocode about
the algorithm in Table~\ref{tabstrong}.
%
%
\begin{table}
\caption{The $\varepsilon$-strong algorithm. It iteratively unveils extra
information about the underlying path.
It outputs the collection of intersection layers $\mathcal{P}=\bigvee
_{j=1}^{2^n}\mathcal{I}_{(j-1)2^{-n},j2^{-n}}$}
\label{tabstrong}
\begin{tabular*}{\textwidth}{@{\extracolsep{\fill}}l@{}}
\hline
\textit{$\varepsilon$-strong($X_0$, $X_1$, $n$):}\\
\quad{1.} Initialize $\mathcal{U}_{0,1}$, $\mathcal
{L}_{0,1}$, set $\mathcal{I}_{0,1}=\{X_0,X_1,\mathcal
{U}_{0,1},\mathcal{L}_{0,1}\}$.
Set $\mathcal{P}=\{\mathcal{I}_{0,1}\}$ and $i=1$.\\
\quad{2.} For each of the $2^{i-1}$ intersection layers in
$\mathcal{P}$, say $\mathcal{I}_{s,t}$, do the following:
\\
\qquad\phantom{i}{i.} Bisect the information $\mathcal{I}_{s,t}$ into
$\mathcal{I}_{s,t^{*}}$, $\mathcal{I}_{t^{*},t}$, where
$t^{*}=(t+s)/2$.
\\
\qquad{ii.} Refine $\mathcal{I}_{s,t^{*}}$, $\mathcal
{I}_{t^{*},t}$ until
the width of their layers is not greater
than $\sqrt{(t-s)/2}$.
\\
\quad{3.} Collect the updated information, $\mathcal
{P}=\bigvee
_{j=1}^{2^i}\mathcal{I}_{(j-1)2^{-i},j2^{-i}}$.
\\
\quad{4.} If $i<n$ set $i=i+1$ and return to Step 2; otherwise
return $\mathcal{P}$.\\
\hline
\end{tabular*}
\end{table}

Utilising the information the $\varepsilon$-strong algorithm returns, we define
the dominating processes as follows:
%
%
\begin{eqnarray}
\label{equplo}
X^{\uparrow}_{u}(n) &=& \sum_{i=1}^{2^n}U^{\uparrow}_{(i-1)2^{-n},i
2^{-n}}\cdot
\mathrm{I}_{ u \in( (i-1)2^{-n},i 2^{-n} ]} ,
\nonumber
\\[-8pt]
\\[-8pt]
\nonumber
X^{\downarrow}_{u}(n) &= &\sum_{i=1}^{2^n}L^{\downarrow
}_{(i-1)2^{-n},i 2^{-n}}\cdot
\mathrm{I}_{ u \in( (i-1)2^{-n},i 2^{-n} ]} .
\end{eqnarray}
The square-root rate at Step 2.ii of the algorithm in Table \ref
{tabstrong} is to guarantee
convergence of the dominating paths with minimal computing cost: it provides
the correct distribution of effort between time-interval and extrema-interval
bisections. To understand this, note that the range of a Brownian
motion (or a Brownian bridge)
on $[0,2^{-n}]$ scales as $\mathcal{O}(2^{-n/2})$; see, for instance,
\cite{revu99}.
Thus, had we used the actual Brownian minima and maxima to define
dominating processes for the
Brownian path in the way of (\ref{equplo}) the rate of convergence
would have been $\mathcal{O}(2^{-n/2})$;\vadjust{\goodbreak}
we cannot exceed such a rate, but we can preserve it if our extrema are
not further than $\mathcal{O}(2^{-n/2})$
from the actual ones. This intuitive statement will be made rigorous in
the sequel, when an explicit
result on the rate of convergence of the dominating processes in
$L_1$-norm is given.

%
\begin{figure}[t!]

\includegraphics{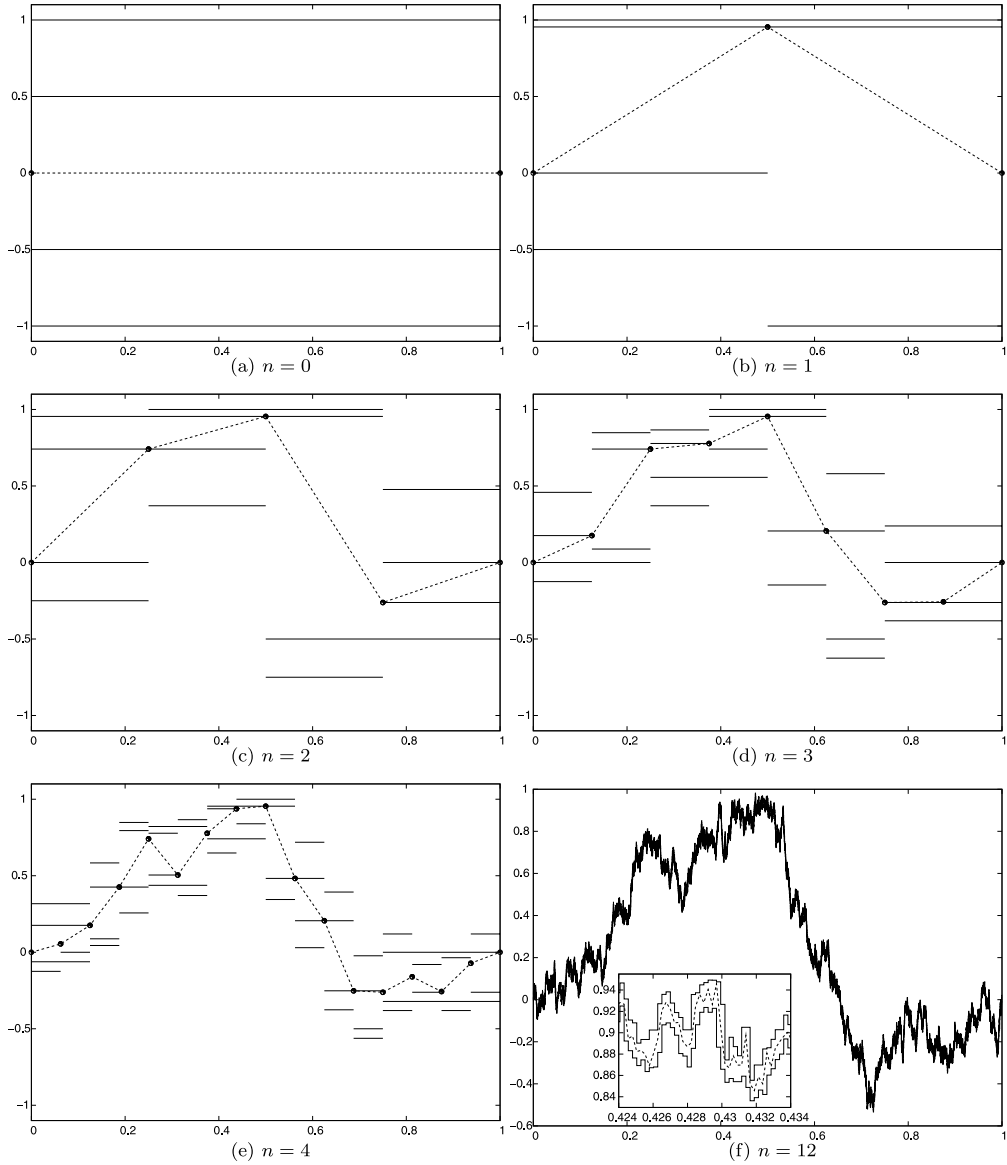}

\caption{The $\varepsilon$-strong algorithm as applied on a personal
computer. For each step $n$,
the horizontal black lines show the allowed interval for the minima and the maxima:
this information is separately
available for all $2^n$ time sub-intervals partitioning $[0,1]$. Note
that the last graph corresponds
to $n=12$, with the subplot in its frame corresponding to a zooming on
the position of the paths
on the time interval $[0.424,0.434]$.}
\label{figstrong}
\end{figure}

Figure~\ref{figstrong} shows successive steps of the $\varepsilon
$-strong algorithm as implemented
on a computer. For each $n$, the horizontal black lines show the interval where the
maxima and the minima
are located: this information is available for all $2^n$ sub-intervals
bisecting the initial time interval $[0,1]$. The dashed black line corresponds to
the linear interpolation of successively unveiled positions of the
underlying Brownian path.
The last graph (f) corresponds to $n=12$; in this case, we have zoomed
on a particular subinterval of $[0,1]$ to
be able to visualise the difference between the bounding paths and the
underlying Brownian one.
%
\subsection{Convergence properties}
Almost sure convergence of the dominating paths follows directly from the
continuity of the Brownian path $X$. The analytical proof is given in
the following proposition.
\begin{prop}
Consider the continuous-time processes $X^{\uparrow}(n)$,
$X^{\downarrow}(n)$
defined in (\ref{equplo}). Then, the convergence in supremum norm in
(\ref{eqconverge}) will hold in the limit
$n\rightarrow\infty$.
\end{prop}
\begin{pf}
For a Brownian bridge $X$ on $[0,1]$, we consider:
\[
D_n := \sup_{1\le i\le2^n}\bigl(M_{(i-1)2^{-n},i 2^{-n}}-m_{(i-1)2^{-n},i
2^{-n}}\bigr) .
\]
Uniform continuity implies that, with probability $1$:
\[
\lim_{n\rightarrow\infty} D_n = 0 .
\]
Now, we have that:
\[
\sup_{u\in[0,1]}|X^{\uparrow}_{u}(n)-X^{\downarrow}_{u}(n)| \le
D_n + 2\cdot2^{-n/2} \rightarrow0 ,
\]
where we have used the fact that Step 2.ii of the $\varepsilon$-strong
algorithm guarantees
that
\begin{eqnarray*}
U^{\uparrow}_{(i-1)2^{-n},i 2^{-n}}&\le& M_{(i-1)2^{-n},i 2^{-n}} +
2^{-n/2} ,
\\
L^{\downarrow}_{(i-1)2^{-n},i 2^{-n}}&\ge& m_{(i-1)2^{-n},i 2^{-n}} -
2^{-n/2} .
\end{eqnarray*}
\upqed\end{pf}

A more involved result can give the rate of convergence of
the dominating processes
and will be of practical significance for the efficiency of Monte Carlo
methods based on
the $\varepsilon$-strong algorithm.
\begin{prop}
\label{prl1}
Consider the $L_1$-distance:
\[
|X^{\uparrow}(n)-X^{\downarrow}(n)|_1 = \int
_{0}^{1}|X^{\uparrow}_{u}(n)-X^{\downarrow}_{u}(n)|\,\mathrm{d}u .\vadjust{\goodbreak}
\]
Then:
\[
2^{n/2} \times\mathrm{E} [ |X^{\uparrow}(n)-X^{\downarrow
}(n)|_1] =
\mathcal{O}(1) .
\]
\end{prop}
\begin{pf}
We proceed as follows:
\begin{eqnarray}
\label{eqL1limit}
|X^{\uparrow}(n)-X^{\downarrow}(n)|_1 &=& \sum_{i=1}^{2^n}
\bigl(
U^{\uparrow}_{(i-1)2^{-n},i 2^{-n}}-L^{\downarrow}_{(i-1)2^{-n},i 2^{-n}}
\bigr)\cdot2^{-n}
\nonumber
\\[-8pt]
\\[-8pt]
\nonumber
&\le&\sum_{i=1}^{2^n}
\bigl( M_{(i-1)2^{-n},i 2^{-n}}-m_{(i-1)2^{-n},i 2^{-n}} + 2\cdot
2^{-n/2} \bigr)
\cdot2^{-n} ,
\end{eqnarray}
the inequality being a direct consequence of Step 2.ii of
the $\varepsilon$-strong algorithm in Table~\ref{tabstrong}.
Consider now the path from $X_{(i-1)2^{-n}}$ to $X_{i 2^{-n}}$.
Let $Z$ be a Brownian bridge from $Z_{0}=0$ to $Z_{2^{-n}}=0$;
we denote by $M_z$ and $m_{z}$ its maximum and minimum, respectively.
Conditionally on $X_{(i-1)2^{-n}}$ and $X_{i 2^{-n}}$, a known
property of
the Brownian bridge implies (see, e.g.,~\cite{kara91})
that:
\[
X_{t+(i-1)2^{-n}} = Z_t + \biggl(1-\frac{t}{2^{-n}}\biggr) X_{(i-1)2^{-n}} +
\frac{t}{2^{-n}} X_{i 2^{-n}},\qquad
t\in[0,2^{-n}] ,
\]
in the sense that the processes on the two sides of the above equation
have the same distribution.
It is now clear that:
\[
M_{(i-1)2^{-n},i 2^{-n}}-m_{(i-1)2^{-n},i 2^{-n}} \le\bigl|X_{i 2^{-n}}
- X_{(i-1)2^{-n}}\bigr| + (M_z - m_z) .
\]
So, taking expectations at (\ref{eqL1limit}), we get:
\[
\label{eqnew}
\mathrm{E} [ |X^{\uparrow}(n)-X^{\downarrow}(n)|_1
] \le\mathrm{E} [
M_z-m_z ] + 2\cdot2^{-n/2} + \sum_{i=1}^{2^n}
\mathrm{E}\bigl |X_{i 2^{-n}} - X_{(i-1)2^{-n}}\bigr| 2^{-n}
\]
The finite-dimensional distributions of the initial Brownian bridge
from $X_0$ to $X_1$ imply that:
\[
X_{i 2^{-n}} - X_{(i-1)2^{-n}}\sim N\bigl((X_1-X_0)2^{-n},2^{-n}(1-2^{-n})\bigr)
,
\]
which gives directly that:
\[
\mathrm{E} \bigl|X_{i 2^{-n}} - X_{(i-1)2^{-n}}\bigr| = \mathcal{O}(2^{-n/2}) .
\]
It remains to show that $\mathrm{E} [ M_z-m_z ]=\mathcal
{O}(2^{-n/2})$ to complete the proof.
Now, self-similarity of Brownian motion implies that:
\[
Z_u = 2^{-n/2} \tilde{Z}_{u/2^{-n}},
\]
where $\tilde{Z}$ is a Brownian bridge from $\tilde{Z}_0=0$ to
$\tilde{Z}_1=0$.
Let $\tilde{M}_z$, $\tilde{m}_z$ be the maximum and minimum of
$\tilde
{Z}$. Due to the self-similarity,
we have
\[
M_z - m_z = 2^{-n/2} (\tilde{M}_z -\tilde{m}_z) .
\]
Since $\tilde{M}_z -\tilde{m}_z$ in a random variable of finite
expectation (see, e.g.,~\cite{kara91}),
we obtain directly that $\mathrm{E} [ M_z-m_z ]=\mathcal
{O}(2^{-n/2})$ which completes the proof.
\end{pf}

%

%
\section{\texorpdfstring{The $\zeta$-function}{The zeta-function}}
\label{secz}
%
We have yet to present the sampling methods employed when refining or bisecting
an intersection layer during the execution of the $\varepsilon$-strong algorithm,
thus constituting the building blocks of our algorithm.
All probabilities involved in these methods can be expressed in terms
of a hitting
probability of the Brownian path.
We denote by
\[
\mathbb{W}^{(l,x,y)}
\]
the probability law of a Brownian bridge from $X_0=x$ to $X_{l}=y$.
Let $\zeta(L,U;l,x,y)$, with $L<U$, be the probability that the
Brownian bridge
escapes the interval $[L,U]$.
That is:
\[
\zeta(L,U;l,x,y) = \mathbb{W}^{(l,x,y)} [ m_{0,l}<L \mbox{ or }
M_{0,l} > U ] .
\]
We also define:
%
%
\begin{equation}
\label{eqGamma}
\gamma(L,U;l,x,y) = 1 - \zeta(L,U;l,x,y) .
\end{equation}
These probabilities can be calculated analytically in terms of an
infinite series.
The result is based on a partition of Brownian paths w.r.t. to a
trace they leave
on two bounding lines and can be attributed back to~\cite{doob49};
for more recent references see
\cite{ande60,wang01,case08}. We define for $j \geq1$,
\begin{eqnarray}\label{eqanalytical}
\bar{\sigma}_{_{j}}(x,y,\delta,\xi) &= &\exp\biggl\{-\frac{2}{l} [
\delta j+\xi-x ]
[ \delta j+\xi-y ]\biggr\},
\nonumber
\\[-8pt]
\\[-8pt]
\nonumber
\bar{\tau}_{_{j}}(x,y,\delta) &=&
\exp\biggl\{-\frac{2j}{l} [ \delta^2j+\delta(x-y) ]\biggr\} .
\end{eqnarray}
Then, Theorem 3 of~\cite{wang01} yields
%
%
\begin{equation}
\label{eqgamma}
\zeta(L,U;l,x,y) = \cases{
\displaystyle\sum_{j=1}^{\infty}(\sigma_{_{j}}-
\tau_{_{j}}) ,& \quad$L<x,y<U ,$ \vspace*{2pt}\cr
1 ,& \quad$\mbox{otherwise} ,$}
\end{equation}
where
\begin{eqnarray}\label{eqtwo}
\sigma_{j}&=&
\bar{\sigma}_{_{j}}(x,y,U-L,L)+\bar{\sigma}_{_{j}}(-x,-y,U-L,-U)
,
\nonumber
\\[-8pt]
\\[-8pt]
\nonumber\
\tau_{j} &=&
\bar{\tau}_{_{j}}(x,y,U-L)+\bar{\tau}_{_{j}}(-x,-y,U-L) .
\medskip
\end{eqnarray}
The infinite series in (\ref{eqgamma}) exhibits a monotonicity property
which will be exploited by our simulation algorithms.
We consider the sequence $\{S_n\}$, with $S_n=S_n(L,U;l,x,y)$, defined as:
%
%
\begin{equation}
\label{eqSigma}
S_{2n-1} = \sum_{j=1}^{n-1}(\sigma_{j}-\tau_{j}) + \sigma_n,\qquad
S_{2n}=S_{2n-1}-\tau_{n} ,
\end{equation}
when $L<x,y<U$, otherwise $S_n\equiv1$.
Then:
%
%
\begin{equation}
\label{eqtelescopic}
0<S_{2n} \leq S_{2n+2} \le\zeta\le S_{2n+1} \le S_{2n-1}
\end{equation}
for all $n\ge1$; for a proof see~\cite{case08} or~\cite{besk08}.
%
\subsection{\texorpdfstring{$\zeta$-Derived events}{zeta-Derived events}}
We can combine $\zeta$-probabilities to calculate other conditional
probabilities arising in the context of the $\varepsilon$-strong algorithm.
We begin with the following definition:
\[
\beta(L^{\downarrow},L^{\uparrow},U^{\downarrow
},U^{\uparrow};l,x,y):=
\mathbb{W}^{(l,x,y)} [ L^{\downarrow} < m_{0,l} < L^{\uparrow
}, U^{\downarrow} <
M_{0,l}<U^{\uparrow} ] .
\]
Now, we have the set equality:
\begin{eqnarray}\label{eqset}
&&\{ L^{\downarrow} < m_{0,l} < L^{\uparrow}, U^{\downarrow}
< M_{0,l}<U^{\uparrow} \}
\nonumber
\\[-8pt]
\\[-8pt]
\nonumber
&&\quad=
\{ L^{\downarrow} < m_{0,l}, M_{0,l} < U^{\uparrow} \} -
 \{L^{\uparrow} < m_{0,l}, M_{0,l} <U^{\uparrow}\}\cup\{
L^{\downarrow}<m_{0,l},
M_{0,l}<U^{\downarrow} \} .\qquad
\end{eqnarray}
Thus, taking probabilities and recalling the definition of $\gamma$ in
(\ref{eqgamma}), we find that:
\begin{eqnarray}\label{eqbeta}
\beta(L^{\downarrow},L^{\uparrow},U^{\downarrow
},U^{\uparrow}; l,x,y)& =&
\gamma(L^{\downarrow},U^{\uparrow};l,x,y) - \gamma
(L^{\uparrow},U^{\uparrow};l,x,y)
\nonumber
\\[-8pt]
\\[-8pt]
\nonumber
&&{}-\gamma(L^{\downarrow},U^{\downarrow};l,x,y)
+ \gamma(L^{\uparrow},U^{\downarrow};l,x,y) .
\end{eqnarray}
Before the next event, we enrich the notation for
the Brownian bridge measure. We define (for $0<q<l$):
\[
\mathbb{W}^{(l,x,y)}_{(q,w)} [ \cdot] = \mathbb{W}^{(l,x,y)} [
\cdot\mid X_{q}=w ] .
\]
We set $r=l-q$.
Consider now the conditional probability:
\[
\rho(L^{\downarrow},L^{\uparrow},U^{\downarrow
},U^{\uparrow};q,r,x,w,y) = \mathbb
{W}^{(l,x,y)}_{(q,w)}
[ L^{\downarrow} < m_{0,l} < L^{\uparrow}, U^{\downarrow} <
M_{0,l}<U^{\uparrow} ] .
\]
%
%
Using again the set equality (\ref{eqset}), and taking probabilities
under $\mathbb{W}^{(l,x,y)}_{(q,w)}$, we obtain:
%
%
\begin{equation}
\label{eqrho}
\rho(L^{\downarrow},L^{\uparrow},U^{\downarrow
},U^{\uparrow};q,r,x,w,y) =
\gamma_1\gamma_2 - \gamma_3\gamma_4
-\gamma_5 \gamma_6 + \gamma_7\gamma_8,
\end{equation}
where we have defined:
\begin{eqnarray*}
\gamma_1 &=& \gamma(L^{\downarrow},U^{\uparrow};q,x,w) ,\qquad
\gamma_2 = \gamma(L^{\downarrow},U^{\uparrow};r,w,y) ,\qquad
\gamma_3 = \gamma(L^{\uparrow},U^{\uparrow};q,x,w) ,\\
\gamma_4 &=& \gamma(L^{\uparrow},U^{\uparrow};r,w,y) ,\qquad
\gamma_5 = \gamma(L^{\downarrow},U^{\downarrow};q,x,w) ,\qquad
\gamma_6 = \gamma(L^{\downarrow},U^{\downarrow};r,w,y) ,\\
\gamma_7 &=& \gamma(L^{\uparrow},U^{\downarrow};q,x,w) ,\qquad
\gamma_8 = \gamma(L^{\uparrow},U^{\downarrow};r,w,y) .
\end{eqnarray*}
Note that the product terms arise due to the independency of the
Brownian bridges on $[0,q]$ and $[q,l]$.
We will be using these expressions for $\beta(\cdot;\cdot)$ and
$\rho
(\cdot;\cdot)$
in the sequel.
%
\subsection{\texorpdfstring{Simulation of $\zeta$-derived events}{Simulation of zeta-derived events}}
\label{seczeta}
We will need to be able to decide whether events of probability $\zeta$
have occurred or not. In a simulation context,
this corresponds to determining the value of the binary variable
$\mathrm{I}_{ R<\zeta}$
for $R\sim\operatorname{Un} [ 0,1 ]$. With (\ref{eqtelescopic}) in mind,
we define:
\[
J = \inf\{n\ge1\dvtx  n \mbox{ odd, }S_n < R \mbox{ or } n \mbox{ even, }S_n > R \} .
\]
Due to the alternating monotonicity property (\ref{eqtelescopic}) of $S_n$:
\[
\mathrm{I}_{ R<\zeta} = \mathrm{I}_{ J\ \mathrm{is\ even}} .
\]
Thus, we need a.s. finite number of
$J$ computations to evaluate $\mathrm{I}_{ R<\zeta}$. Note that $S_{n}$
converges to its limit exponentially fast, so $J$ will be of small expectation;
one can easily verify that all its moments are finite. Such an approach
was also followed in
\cite{besk08}.

In a more general context, we will also be required to decide
if events of probability $\beta(\cdot;\cdot)$ or $\rho(\cdot;\cdot)$
have taken place or not; we will in fact be considering even more
complex events related with the $\zeta$-function.
In the most encompassing scenario, when executing our sampling methods,
we will be required to compare a given real number $R$ with
$Z(\zeta_{1},\zeta_2,\ldots,\zeta_m)$ for some given function $Z$,
with the different $\zeta_{i}$'s corresponding to different choices of
the arguments $l,x,y,L,U$ for $\zeta( \cdot;\cdot)$.
Using the monotonicity property (\ref{eqtelescopic}), we will be able
to develop corresponding alternating sequences
$S_n^{Z}$ such that:
%
%
\begin{eqnarray}
\label{eqtelescopic-Z}
S^Z_{2n} &\le& S^Z_{2n+2} \le Z(\zeta_1,\zeta_2,\ldots, \zeta_m) \le
S^Z_{2n+1} \le S^Z_{2n-1} ;
\nonumber
\\[-8pt]
\\[-8pt]
\nonumber
\lim_{n\rightarrow\infty} S^Z_{n} &=& Z(\zeta_1,\zeta_2,\ldots,
\zeta_m)
,
\end{eqnarray}
and proceed as above. Analytically, we will determine the value of the
comparison binary indicator $\mathrm{I}_{ R<Z(\zeta_1,\zeta_2,\ldots
,\zeta_m)}$ as follows:

\textit{Calculate $S_n^{Z}$ until the first $n$ such that either $n$ is odd and
$S_n^Z < R$ (whence return $0$)
or $n$ is even and $S_n^Z > R$ (whence return $1$).}

%

%
\section{Distributions and their simulation}
\label{secdetails}
%
We will now describe analytically all simulation algorithms employed at
the development
of the $\varepsilon$-strong algorithm presented in Table~\ref{tabstrong}.
In particular, one has to develop sampling methods to carry out the
refinement and bisection
(see Section~\ref{seclayer})
of the intersection layer $\mathcal{I}_{s,t}$.
To simplify the presentation, when conditioning on $X_s$, $X_{t^{*}}$
or $X_t$
we will make the correspondence:
\begin{eqnarray*}
x&=&X_s ,\qquad w=X_{t^*},\qquad y=X_t , \\
l&=&t-s, \qquad q=t^*-s,\qquad  r= t-t^* .
\end{eqnarray*}
%
\subsection{Bisection of $\mathcal{I}_{s,t}$: Sampling the middle
point $X_{t^*}$}
\label{subsecmiddle}

Bisection of $\mathcal{I}_{s,t}=\{X_s,X_t,\mathcal{L}_{s,t}, \mathcal
{U}_{s,t}\}$, with $\mathcal{L}_{s,t}=[L^{\downarrow},L^{\uparrow}]$,
$\mathcal{U}_{s,t}=[U^{\downarrow},U^{\uparrow}]$,
begins by sampling
a point of the Brownian bridge
conditionally on the collected information about its minimum and maximum;
this is Step 1 of Table~\ref{tabbisect}.
Such a conditional distribution is analytically tractable via Bayes' theorem.
\begin{prop}
\label{prmiddle}
The distribution $\mathbb{W} [ X_{t^{*}}\mid\mathcal{I}_{s,t} ]$,
with $t^*\in[s,t]$,
has probability density:
\[
f(w) \propto\rho(L^{\downarrow},L^{\uparrow},U^{\downarrow
},U^{\uparrow};q,r,x,w,y)
\times\pi(w)
\]
where $\rho(\cdot;\cdot)$ is defined in (\ref{eqrho}) and
\[
\pi(w) = \exp\biggl\{ -\frac{1}{2}\biggl(w-\biggl( \frac{r}{l} x+\frac
{q}{l} y \biggr)\biggr)^2\Big/
\biggl( \frac{qr}{l} \biggr) \biggr\} .
\]
\end{prop}
\begin{pf}
The function $\pi(w)$ corresponds to the prior (unnormalised) density
for the middle point $X_t^{*}\vert X_s, X_t$ which is easily found
to be normally distributed with mean and variance as
implied by the expression for $\pi(w)$.
So, following the definition of $\rho(\cdot;\cdot)$ in (\ref{eqrho}),
the stated result is an application of Bayes' theorem.
\end{pf}

We will develop a method for sampling from $f(w)$.
It is easy to construct an alternating series bounding $f(w)$.
Let:
\[
\zeta_i=1-\gamma_i ,\qquad 1\le i \le8 ,
\]
for the eight $\gamma$-functions appearing at the definition of $\rho$
in (\ref{eqrho}).
%
%
Let $\{S_{i,n}\}_{n\ge1}$ be the alternating series (\ref
{eqtelescopic}) for $\zeta_i$,
for each $1\le i \le8$; that is:
%
%
\begin{equation}
\label{eqtel}
0<S_{i,2n} \leq S_{i,2n+2} \le\zeta_{i} \le S_{i,2n+1} \le
S_{i,2n-1} ,
\end{equation}
with $\lim_{n\rightarrow\infty} S_{i,n}= \zeta_i$.
Consider the sequence $\{S_{n}^{Z}\}$ defined as follows:
\begin{eqnarray} \label{eqenvelop}
S_{n}^{Z} &=& (1-S_{1,n+1}-S_{2,n+1}+S_{1,n}S_{2,n}) - (
1-S_{3,n}-S_{4,n}+S_{3,n+1}S_{4,n+1})
\nonumber
\\[-8pt]
\\[-8pt]
\nonumber
&&{}-( 1-S_{5,n}-S_{5,n}+S_{5,n+1}S_{6,n+1}) + (1-S_{7,n+1}-S_{8,n+1}+
S_{7,n}S_{8,n}) .
\end{eqnarray}
Due to (\ref{eqtel}), one can easily verify that $\{S_n^Z\}$ is an
alternating sequence for
$\rho(\cdot;\cdot)$, in the sense that:
%
%
\begin{equation}
\label{eqtelrho}
S^{Z}_{2n} \leq S^{Z}_{2n+2} \le\rho(L^{\downarrow},L^{\uparrow
},U^{\downarrow},U^{\uparrow},q,r,x,w,y) \le S^{Z}_{2n+1}
\le S^{Z}_{2n-1}
\end{equation}
with $\lim_{n\rightarrow\infty} S^{Z}_{n} = \rho(L^{\downarrow
},L^{\uparrow},U^{\downarrow},U^{\uparrow};q,r,x,w,y)$.

We exploit this structure to build a rejection sampler
to draw from the density $f(w)$ in Proposition~\ref{prmiddle}.
We will use proposals from:
\[
f_{2n+1}(w) = S_{2n+1}^{Z}(w)\times\pi(w) ,
\]
where we have emphasized the dependence of $S^{Z}_{2n+1}$ on the
argument $w$.
Note that the domain of both $f(w)$, $f_{2n+1}(w)$ is $[L^{\downarrow
},U^{\uparrow}]$.
Now, we will illustrate that $S^{Z}_{2n+1}(w)$ has a concrete structure
that we will exploit for our sampler.
Consider the first of the four terms forming up $S^{Z}_{2n+1}$ from~(\ref{eqenvelop}):
%
%
\begin{equation}
\label{eqterm}
1-S_{1,2n+2}-S_{2,2n+2}+S_{1,2n+1}S_{2,2n+1} .
\end{equation}
Following the analytical definition of the alternating sequences in
equations (\ref{eqanalytical}), (\ref{eqtwo}), (\ref{eqSigma}), both
$S_{1,n}$ and $S_{2,n}$, can be expressed as a sum of $2n$ terms each
having the exponential structure
$\pm\exp\{a + b w\} \mathrm{I}_{ L^{\downarrow}<w<U^{\uparrow
}}$ for appropriate
constants $a,b$ varying among the $2n$ terms. Thus, the quantity in
(\ref{eqterm}) can be expressed as:
\[
1+\sum_{i=1}^{k_{1,n}}(-1)^{c_i}\exp\{a_i + b_i w\} \mathrm{I}_{
L^{\downarrow}<w<U^{\uparrow}}
\]
for $k_{1,n} = 4\{(2n+1)^2+(2n+2)\}$, and constants $a_i$, $b_i$, $c_i$
with $c_i\in\{0,1\}$.
Working similarly for all four summands forming up $S^{Z}_{2n+1}$ in
(\ref{eqenvelop}), we get that the function
$f_{2n+1}(w)$ can in fact be written as the weighted sum:
%
%
\begin{equation}
\label{eqenvelope}
f_{2n+1}(w) = \sum_{i=1}^{k_n}(-1)^{c_i}\exp\{a_i + b_i w\} \mathrm
{I}_{ L_i<w<U_i}\times\pi(w)
\end{equation}
for $k_n = 2(k_{1,n}+k_{2,n})$ with $k_{2,n} = 4\{(2n+2)^2+(2n+1)\}$, and
some explicit constants $a_i$, $b_i$, $c_i\in\{0,1\}$, $L_i$, $U_i$.
Experimentation has showed that $f_1$ is already a very good envelope function
for the rejection sampler, in which case $k_n\equiv k_0 = 64$; this is
not accidental, and relates with
the rapid exponential convergence of the alternating sequence in (\ref
{eqSigma}) to its limit.
The cdf, say $F_{1}(w)$, corresponding to the unormalised density
function $f_{1}(w)$ can be analytically identified since
integrals for each of the summands in (\ref{eqenvelope}) can be
expressed as differences of the cdf of the standard Gaussian
distribution. Samples from $f_{1}(w)$
can then be generated using the inverse cdf method, that is, by
returning $F_{1}^{-1}(R)$ for $R\sim\operatorname{Un}[0,1]$.
$F_1^{-1}$ cannot be found analytically, but numerical methods can
return $F_{1}^{-1}(R)$,
up to maximum allowed computer accuracy, exponentially fast. We have
used \texttt{MATHEMATICA} to
automatically calculate
all integrals giving the cdf, and then incorporated the calculation
into a \texttt{C++} code.

Summarising, our rejection sampler will be as described below, where
for simplicity we write
$\rho(w) \equiv\rho(L^{\downarrow},L^{\uparrow
},U^{\downarrow},U^{\uparrow};q,r,x,w,y)$:
\begin{description}
\item\textit{Repeat until the first accepted draw:}

\item\textit{Propose $w\sim f_1$ and accept with probability $f(w)/f_1(w)\equiv\rho
(w)/S^{Z}_{1}(w)$.}
\end{description}

%

%

\noindent Note here that the acceptance probability involves $\rho(w)$
which is made up of
eight infinite series, see (\ref{eqrho}). We avoid approximations by
using the alternating
construction (\ref{eqtelrho}) and employ the methods of Section \ref
{seczeta} to obtain
the value of the decision indicator $\mathrm{I}_{ R< \rho
(w)/S^{Z}_{1}(w)}$ for some
$R\sim\operatorname{Un} [ 0,1 ]$.

As shown in Step 1 of Table~\ref{tabbisect}, once $X_{t^*}$ is
obtained, we adjust the allowed range
for the extrema of the bridge on $[s,t]$ by simply setting
$U^{\downarrow} = U^{\downarrow}\vee X_{t^*}$, $L^{\uparrow
} = L^{\uparrow}\wedge X_{t^*}$.
%
\subsection{Bisection of $\mathcal{I}_{s,t}$: Updating the Layers
given $X_{t^{*}}$}
\label{secE}

At the second step of the bisection procedure, see Table~\ref{tabbisect},
we obtain separate information
for the extrema of the two newly formed bridges
given the middle point $X_{t^{*}}$:
the one bridge being from $X_s$ to $X_{t^*}$, the
other from $X_{t^*}$ to $X_t$.
In particular, the algorithm will decide over the range of the four
newly formed layers, $\mathcal{L}_{s,t^{*}}$,
$\mathcal{U}_{s,t^*}$, $\mathcal{L}_{t^*,t}$, $\mathcal{U}_{t^*,t}$
in the following manner:
for the case of $\mathcal{L}_{s,t^{*}}$
for instance a decision will be made over whether $m_{s,t^*}$ lies in
$[L^{\downarrow},L^{\uparrow}]$ (which is the allowed range for
the minimum of the
original bridge on $[s,t]$) or in $[L^{\uparrow},X_{s}\wedge
X_{t^*}]$. The
apparent analogues apply in the case of the three other layers.

One might initially think that there are in total $2^4$ different
scenaria for the four layers.
But one has to remember that the update has to respect the information
in $\mathcal{I}_{s,t}$, so that at least one of the two minima (resp.
maxima) on $[s,t^*]$ and $[t^*,t]$ \textit{must} lie in
$[L^{\downarrow},L^{\uparrow}]$
(resp. $[U^{\downarrow},U^{\uparrow}]$). In particular, there
are in fact nine
different possible scenaria,
which are the ones shown in Table~\ref{tupdate} (labelled as events $\{
E=i\}$, for $1\le i\le9$):
a value of $1$ in Table~\ref{tupdate} means that the corresponding
minimum or maximum will still be found
within the allowed range for the original bridge on $[s,t]$, whereas a
value of $0$ means that
the second option occurs and the extremum will be shifted inwards. For instance,
a value of $0$ for the indicator variable concerning $m_{s,t^*}$,
$M_{s,t^*}$, $m_{t^*,t}$ or $M_{t^*,t}$
implies that $m_{s,t^*}\in[L^{\uparrow},X_s\wedge X_{t^*}]$,
$M_{s,t^*}\in[X_s\vee X_{t^*},U^{\downarrow}]$,
$m_{t^*,t}\in[L^{\uparrow},X_{t^*}\wedge X_{t}]$ or
$M_{t^*,t}\in[X_{t^*}\vee X_{t},U^{\downarrow}]$, respectively.
%

%
%
\begin{table}
\caption{The nine possible scenaria for the extrema of the two
Brownian bridges
(from $X_s$ to $X_t^{*}$ and from $X_{t^*}$ to $X_t$)}
\label{tupdate}
\begin{tabular*}{\textwidth}{@{\extracolsep{\fill}}lllll@{}}
\hline
 & \multicolumn{2}{l}{Left bridge} & \multicolumn{2}{l@{}}{Right bridge}
\\[-6pt]
Event & \multicolumn{2}{c}{\hrulefill} & \multicolumn{2}{c@{}}{\hrulefill} \\
$E=i$ & $\mathrm{I}_{ m_{s,t^*}\in[L^{\downarrow}, L^{\uparrow
}]}$ &
$\mathrm{I}_{ M_{s,t^*}\in[U^{\downarrow},U^{\uparrow}]}$ &
$\mathrm{I}_{ m_{t^*,t}\in[L^{\downarrow}, L^{\uparrow}]}$ &
$\mathrm{I}_{ M_{t,t^*}\in[U^{\downarrow},U^{\uparrow}]}$ \\
\hline
$i=1$ & $1$ & $1$
& $1$ & $1$ \\
$i=2$ & $1$ & $1$
& $0$ & $1$ \\
$i=3$ & $1$ & $1$
& $1$ & $0$\\
$i=4$ & $1$ & $1$
& $0$ & $0$ \\
$i=5$ & $0$ & $1$
& $1$ & $1$ \\
$i=6$ & $0$ & $1$
& $1$ & $0$ \\
$i=7$ & $1$ & $0$
& $1$ & $1$ \\
$i=8$ & $1$ & $0$
& $0$ & $1$ \\
$i=9$ & $0$ & $0$
& $1$ & $1$ \\
\hline
\end{tabular*}
\end{table}
%
%
\begin{table}[b]
\tablewidth=285pt
\caption{The conditional probabilities for each of the events in Table~\protect\ref{tupdate}}
\label{tcalculate}
\begin{tabular*}{285pt}{@{\extracolsep{\fill}}ll@{}}
\hline
$i$ & $\mathrm{P} [ E=i | \mathcal{I}_{s,t},X_{t^*} ]\times\rho
(L^{\downarrow},L^{\uparrow},U^{\downarrow},U^{\uparrow
};q,r,x,w,y)$ \\
\hline
1 & $\beta(L^{\downarrow}, L^{\uparrow}, U^{\downarrow
},U^{\uparrow};q,x,y)\times
\beta(L^{\downarrow}, L^{\uparrow}, U^{\downarrow
},U^{\uparrow};r,w,y ) $ \\
2 & $\beta(L^{\downarrow}, L^{\uparrow}, U^{\downarrow
},U^{\uparrow};q,x,w)\times
\beta(L^{\uparrow}, w_y, U^{\downarrow},U^{\uparrow};r,w,y
) $\\
3 & $\beta(L^{\downarrow}, L^{\uparrow}, U^{\downarrow
},U^{\uparrow};q,x,w )\times
\beta(L^{\downarrow}, L^{\uparrow}, w^y,U^{\downarrow
};r,w,y ) $\\
4 & $\beta(L^{\downarrow}, L^{\uparrow}, U^{\downarrow
},U^{\uparrow};q,x,w )\times
\beta(L^{\uparrow}, w_y, w^y,U^{\downarrow};r,w,y ) $ \\
5 & $\beta(L^{\uparrow}, w_x, U^{\downarrow},U^{\uparrow
};q,x,w )\times
\beta(L^{\downarrow}, L^{\uparrow}, U^{\downarrow
},U^{\uparrow};r,w,y ) $ \\
6 & $\beta(L^{\downarrow}, L^{\uparrow}, U^{\downarrow
},U^{\uparrow};q,x,w )\times
\beta(L^{\downarrow}, L^{\uparrow}, U^{\downarrow
},U^{\uparrow};r,w,y ) $ \\
7 & $\beta(L^{\uparrow}, w_x, U^{\downarrow},U^{\uparrow
};q,x,w )\times
\beta(L^{\downarrow}, L^{\uparrow}, w^y, U^{\downarrow
};r,w,y ) $ \\
8 & $\beta(L^{\downarrow}, L^{\uparrow}, w^x, U^{\downarrow
};q,x,w )\times
\beta(L^{\uparrow}, w_y, U^{\downarrow},U^{\uparrow};r,w,y
) $ \\
9 & $\beta(L^{\uparrow}, w_x, w^x, U^{\downarrow};q,x,w
)\times
\beta(L^{\downarrow}, L^{\uparrow}, U^{\downarrow
},U^{\uparrow};r,w,y ) $ \\
\hline
\end{tabular*}
\end{table}

The probability for each of the events in Table~\ref{tupdate} can be derived
via functions $\beta(\cdot;\cdot)$ and $\rho(\cdot;\cdot)$
defined in
(\ref{eqbeta}) and (\ref{eqrho}),
respectively.
Recall that we are conditioning upon $\mathcal{I}_{s,t}$ and
$X_{t^{*}}$, so
we work as follows:
\begin{eqnarray*}
\mathrm{P} [ E=i \vert \mathcal{I}_{s,t},X_{t^*} ]&=&
\mathrm{P} [ E=i \vert  m\in[L^{\downarrow},L^{\uparrow}], M\in
[U^{\downarrow},U^{\uparrow}], X_{s}, X_{t^*}, X_t ] \\
&=& \frac
{ \mathrm{P} [ E=i, m\in[L^{\downarrow},L^{\uparrow}], M\in
[U^{\downarrow},U^{\uparrow}]
\vert X_{s},
X_{t^*}, X_t ] }
{ \mathrm{P} [ m\in[L^{\downarrow},L^{\uparrow}], M\in
[U^{\downarrow},U^{\uparrow}] \vert
X_{s}, X_{t^*}, X_t ] }\\
&=& \frac{\mathrm{P} [ E=i | X_{s},
X_{t^*}, X_t ]}{\rho(L^{\downarrow},L^{\uparrow},U^{\downarrow
},U^{\uparrow};q,r,x,w,y)} .
\end{eqnarray*}
Now, conditionally on $\{X_{s}, X_{t^*}, X_t\}$ the law of the path
factorises into two independent Brownian
bridges. Thus, recalling also the definition of $\beta(\cdot;\cdot)$ in
(\ref{eqbeta}), the probability $\mathrm{P} [ E=i \vert X_{s},
X_{t^*}, X_t ]$ in the numerator above can be written as a product of
two $\beta(\cdot;\cdot)$ functions.
The analytical calculation of the numerator, or equivalently of the
product $\mathrm{P} [ E=i \vert \mathcal{I}_{s,t},X_{t^*} ]\times\rho
(L^{\downarrow},L^{\uparrow},U^{\downarrow},U^{\uparrow
};q,r,x,w,y)$,
is given in Table~\ref{tcalculate} where,
to simplify the presentation, we have set:
\[
w_x = x\wedge w,\qquad w^x = x\vee w,\qquad w_y = w\wedge y,\qquad w^y =
w\vee y .
\]

The method to simulate the discrete random variable $E$ could follow
the alternating series approach of Section~\ref{seczeta}.
Analytically, consider the cumulative probability values $p_i=\mathrm
{P} [ E\le i | \mathcal{I}_{s,t},X_{t^*}]$.
A simple inverse cdf sampling method requires finding the index $\inf\{
i\ge1\dvtx  R < p_i\}$ for a $R\sim\operatorname{Un} [0,1]$. Note now that the $p_i$'s
can be bounded above and below by monotone
converging sequences as in (\ref{eqtelescopic-Z}), thus each
comparison $\{R<p_i\}$ can be
carried out via the alternating series approach of Section \ref
{seczeta} without any need for approximations.\vspace*{-2pt}

\subsection{Remaining sampling procedures}\vspace*{-2pt}
A sampling algorithm is required for the refinement of the uncertainty
over the extrema of a Brownian
bridge. As described in Section~\ref{secrefine}, given the current
intersection layer information
$\mathcal{I}_{s,t}$ and in particular the fact
that $M_{s,t}\in[L^{\downarrow},L^{\uparrow}]$, the algorithm
will need to decide
whether the maximum $M_{s,t}$ lies in
$[U^{\downarrow},U^*]$ or in $[U^*,U^{\uparrow}]$, for
$U^*=(U^{\downarrow}+U^{\uparrow})/2$.
Recalling the definition of $\beta(\cdot;\cdot)$ from (\ref{eqbeta}),
it is easy to check that the ratio:
\[
\frac{\beta(L^{\downarrow},L^{\uparrow},U^{*},U^{\uparrow
};l,x,y)}{\beta(L^{\downarrow},L^{\uparrow},U^{\downarrow
},U^{\uparrow};l,x,y)}
\]
provides precisely the probability of the event $\{M_{s,t}\in
[U^{*},U^{\uparrow}] | \mathcal{I}_{s,t}\}$.
Thus, we can again use the alternating sequence construction of Section
\ref{seczeta} to simulate,
without approximation, the binary variable $\mathrm{I}_{ M_{s,t}\in
[U^{*},U^{\uparrow}]}$.
The same approach can be followed for refining the allowed range for
the minimum $m_{s,t}$.

We should also give some details over the initialization of the layers
$\mathcal{U}_{0,1}$ and
$\mathcal{L}_{0,1}$ at the first step of the $\varepsilon$-strong
algorithm in
Table~\ref{tabstrong}
given $X_0$ and $X_1$. (Note that sometimes, as in the example
applications that we consider in the following section,
this initialization steps might not even be necessary, as the problem
at hand provides a natural definition
of $\mathcal{U}_{0,1}$ and $\mathcal{L}_{0,1}$.) One way to proceed
is by specifying increasing
sequences $\{a_i\}_{i\ge0}$, $\{b_i\}_{i\ge0}$, with $a_0=b_0=0$,
growing to $\infty$ and a bivariate index $I$ such that:
\[
\{I=(i,j)\} = \{\bar{x}-a_{i}<m\le\bar{x} - a_{i-1}, \bar{y} +
b_{j-1}<M\le\bar{y} + b_{j}\} ,
\]
where $\bar{x}=x\wedge y$, $\bar{y}=x\vee y$. We can easily identify
the probability distribution
of $I$ under the Brownian bridge dynamics since:
\[
\mathbb{W}^{(1,x,y)} [ I=(i,j) ] =
\beta(\bar{x}-a_{i},\bar{x} - a_{i-1}, \bar{y} + b_{j-1},\bar{y} +
b_{j};1,x,y) .
\]
Thus, we can work as in the case of the simulation of the discrete
variable $E$ in Section~\ref{secE}:
assuming $\tilde{I}=1,2,\ldots$ is some chosen ordering of the states
of $I$, an inverse cdf method
would required finding $\inf\{i\ge1\dvtx R<\mathrm{P} [ I\le i ]\}$ for
$R\sim\operatorname{Un}[0,1]$,
and approximations at the comparison between $R$ and $\mathrm{P} [
I\le i ]$
can be avoided via the alternating series approach. In practice, one
could select some big enough
values for the first couple of elements of the sequences $\{a_i\}$ and
$\{b_i\}$ so that almost all
probability mass is concentrated on $\{I=(i,j)\}$ for $i,j\le2$, and
not a lot of computational
resources are spent on this step.\vspace*{-2pt}


\section{Application: Unbiased estimation of path expectations}\vspace*{-2pt}
\label{secexpaths}
The information provided by the $\varepsilon$-strong algorithm can be exploited
to deliver \textit{unbiased} estimators for path expectations arising in
applications,
avoiding discretization errors characterising standard\vadjust{\goodbreak} approaches.
We emphasize that we mean to sketch here only a potential direction for
application
of the algorithm.
Analytically, consider a nonnegative path functional $F\dvtx C([0,1],\mathbb
{R})\mapsto\mathbb{R}^{+}$
and the expectation:
$\mathrm{E}[ F(X) ]$,
$X$ being a Brownian motion on $[0,1]$. One can easily check, by
integrating out $E$, that:
%
%
\begin{equation}
\label{eqtrick}
\mathrm{I}_{ F(X)>E }\cdot \mathrm{e}^{E} ,\qquad E\sim\mathrm{Exp}(1) ,
\end{equation}
with $E$ being independent of $X$, is an unbiased estimator of $\mathrm{E}[
F(X) ]$.
The $\varepsilon$-strong algorithm could be utilised here to unbiasedly
obtain the value of the binary variable
$\mathrm{I}_{ F(X)>E }$ in finite computations.
We can easily find the second moment of the unbiased estimator in (\ref
{eqtrick}):
%
%
\begin{equation}
\label{eqvariance}
\mathrm{E}\bigl[ \mathrm{I}_{ F(X)>E }\cdot \mathrm{e}^{2E} \bigr] = \mathrm{E}\bigl[
\mathrm{e}^{F(X)} \bigr]-1
.
\end{equation}
We describe for a moment in more detail the identification of $\mathrm
{I}_{F(X)>U}$ via the $\varepsilon$-strong
algorithm.
Utilising the lower and upper convergent processes $X^{\downarrow
}(n)$, $X^{\uparrow}(n)$ in (\ref{equplo})
one could in many cases analytically identify quantities
$F^{\downarrow}_{n}$, $F^{\uparrow}_{n}$ (realisable with finite computations)
such that:
\begin{eqnarray*}
& \displaystyle F^{\downarrow}_{n} \le F^{\downarrow}_{n+1} \le F(X) \le F^{\uparrow
}_{n+1} \le F^{\uparrow}_{n} ;&\\
&\displaystyle F^{\uparrow}_{n} - F^{\downarrow}_{n} \rightarrow0 .&
\end{eqnarray*}
Given enough iterations, there will be agreement; for the a.s. finite
random instance:
%
%
\begin{equation}
\label{eqkappa}
\kappa= \inf\{n\ge0\dvt \mathrm{I}_{ F^{\downarrow}_{n}>E }=\mathrm
{I}_{ F^{\uparrow}_{n}>E } \}
\end{equation}
we will have
%
%
\begin{equation}
\label{eqtrick1}
\mathrm{I}_{ F(X)>E } = \mathrm{I}_{ F^{\downarrow}_{\kappa}>E } .
\end{equation}
Thus, combining (\ref{eqtrick}) with (\ref{eqtrick1}), we have
developed an unbiased
estimator of a path expectation, involving finite computations.
Certainly, the numerical efficiency of such an estimation will rely
heavily on
the stochastic properties of $\kappa$ and the cost of generating
$F^{\downarrow}_{n}$, $F^{\uparrow}_{n}$,
and of course the variance of the estimator.

The particular derivation of the above unbiased estimator of the path
expectation is by no means restrictive;
one can generate unbiased estimators using distributions other than the
exponential.
Consider the following scenario.
We can generate some preliminary bounds $F^{\downarrow}_{n_0}$,
$F^{\uparrow}_{n_0}$ up to
some fixed or random (depending on $X$) instance $n_0$.
Now, one can easily check (by considering the conditional expectation
w.r.t. $R|X$) that:
%
%
\begin{equation}
\label{eqimproved}
\mathrm{I}_{F(X)>R} F^{\uparrow}_{n_0} + \mathrm{I}_{F(X)<R}
F^{\downarrow}_{n_0};\qquad
R\sim\operatorname{Un}[F^{\downarrow}_{n_0},F^{\uparrow}_{n_0}] ,
\end{equation}
is also an unbiased estimator of $\mathrm{E}[ F(X) ]$.
We have empirically found the estimator (\ref{eqimproved}) to be much
more robust than (\ref{eqtrick}) in
the numerical applications we present in the sequel. This is not
accidental: for instance, considering a random $n_0$
such that $F^{\uparrow}_{n_0}-F^{\downarrow}_{n_0}<C$, for a constant
$C>0$, we get that
the second moment of the estimator (\ref{eqimproved}) will be:
\begin{eqnarray*}
&&\mathrm{E}\bigl[ \mathrm{I}_{F(X)>R} (F^{\uparrow}_{n_0})^2 + \mathrm
{I}_{F(X)<R} (F^{\downarrow}_{n_0})^2 \bigr]  \\
&&\qquad=\mathrm{E}[ F(X)(F^{\uparrow}_{n_0}+F^{\downarrow}_{n_0}) ]-\mathrm
{E}[ F^{\uparrow}_{n_0}F^{\downarrow}_{n_0} ] \le
\mathrm{E}[ F^{2}(X) ] + C \mathrm{E}[ F(X) ]
\end{eqnarray*}
which has now a quadratic structure -- compare this with (\ref
{eqvariance}). In general, increasing
$n_0$ adds to the computational cost per sample, but decreases the variance.
We have empirically found that moderate values of $n_0$ deliver
significantly better estimates than (\ref{eqtrick}) and will be using
such an approach for
our numerical examples in the sequel.


\subsection{Numerical illustrations}
\label{subsecnum}
We will apply the $\varepsilon$-strong algorithm to unbiasedly estimate
some option prices arising in finance.
In particular, option prices are expressed as expectations:
\[
\mathrm{E}[ F(S) ]
\]
of a functional $F(\cdot)$ of the path process $S=\{S_t\}$ modelling
the underlying asset. We will
consider some double-barrier options corresponding the expectations of
the functionals:
\begin{eqnarray}
F_a(S) &=& \mathrm{e}^{-rT}( \sup S_t - K_S )^{+} \mathrm{I}_{ L_S < \inf
S_t < \sup S_t < U_S} ;\label{eqF1}
\\
F_b(S) &=& \mathrm{e}^{-rT}\biggl( \frac{1}{T}\int_{0}^{T}S_t \,\mathrm{d}t - K_S
\biggr)^{+}
\mathrm{I}_{ L_S < \inf S_t < \sup S_t < U_S} ; \label{eqF2} \\
F_c(S) &=& \mathrm{e}^{-rT}( \sup S_t - K_S )^{+} \mathrm{I}_{ L_S < \inf
S_t \mathrm{e}^{-rt} < \sup S_t \mathrm{e}^{-rt} < U_S}
\label{eqF3}
\end{eqnarray}
(where for $x\in\mathbb{R}$, $x^{+}:=x\vee0$) for underlying asset
$S=\{
S_t;t\in[0,T]\}$ modelled via a geometric Brownian motion (we consider
a Black and Scholes framework)
determined as:
%
%
\begin{equation}
\label{eqgeom}
\log S_t = \log S_0 + \biggl(r-\frac{\sigma^2}{2}\biggr) t + \sigma W_t
\end{equation}
for constants $r$ (interest rate), $\sigma>0$ (volatility) and a
Brownian motion $\{W_t\}$.
Also, $T$ above is the maturity time, $K_S$ the strike price and $L_S$,
$U_S$ the lower and upper barriers respectively;
suprema and infima are considered over the time period $[0,T]$.
Note that $\mathrm{E}[ F_b(S) ]$ corresponds to the price of the Asian
option, see, for example,~\cite{roge95}.
The process $S_t$ is an 1--1 transformation of a Brownian motion with
drift. In particular, we can rewrite
the functionals
(\ref{eqF1})--(\ref{eqF3}) as follows:
\begin{eqnarray}
F_a(X) &=& \mathrm{e}^{-rT}( \mathrm{e}^{\sigma\sup X_t} - K_S )^{+} \mathrm{I}_{ L
< \inf X_t < \sup X_t < U} ;\label{eqf1}\\
F_b(X) &=& \mathrm{e}^{-rT}\biggl( \frac{1}{T}\int_{0}^{T}\mathrm{e}^{\sigma X_t} \mathrm{d}t -
K_S \biggr)^{+} \mathrm{I}_{ L < \inf X_t < \sup X_t < U} ; \label
{eqf2}\\
F_c(X) &=& \mathrm{e}^{-rT}\bigl( \mathrm{e}^{\sigma\sup(({r}/{\sigma}) t + X_t)} - K_S
\bigr)^{+}
\mathrm{I}_{ L < \inf X_t < \sup X_t < U} ,
\label{eqf3}
\end{eqnarray}
for $L=\log(L_S)/\sigma$, $U=\log(U_S)/\sigma$, and:
\begin{eqnarray*}
\mathrm{Case}\ F_a,F_b\dvtx&&\qquad  X_t = \log(S_0)/\sigma+\biggl (\frac
{r}{\sigma}-\frac{\sigma}{2}\biggr)t + W_t ; \\
\mathrm{Case}\ F_c  \dvtx&&\qquad  X_t = \log(S_0)/\sigma-\frac{\sigma
}{2} t + W_t .
\end{eqnarray*}
Conditionally on its ending point, the dynamics of the drifted Brownian motion
do not depend on the value of the drift and
coincide with those of a simple Brownian bridge; this is a simple by-product
of the Girsanov theorem, see, for example,~\cite{okse03}.
Thus, the $\varepsilon$-strong algorithm can now deliver convergent,
lower and upper dominating processes for~$X$.

The choices of functionals in (\ref{eqF1})--(\ref{eqF3}) is not accidental:
some generic characteristics of the structure
of each functional (relevant also for other applications) will effect
the set-up of the $\varepsilon$-strong algorithm and its efficiency;
we will say more on this in the sequel.

For all three examples, our general methodology is as follows:
we begin by sampling $X_T$ and, then, the indicator variable
$\mathrm{I}_{ L < \inf X_t < \sup X_t < U}$; if the latter is $0$ the
sample for our unbiased estimator is
simply $0$, otherwise we proceed with applying the methods of the
$\varepsilon$-strong algorithm by initializing
the first intersection layer as $\mathcal{I}_{0,T}=\{X_0,X_T,\mathcal
{U}_{0,T},\mathcal{L}_{0,T}\}$ for intervals $\mathcal
{U}_{0,T}=[X_0\vee X_T,U]$ and $\mathcal{L}_{0,T}=[L,X_0\wedge X_T]$.
In some cases we might not need all of the
machinery of the
$\varepsilon$-strong algorithm to construct the sequences $F^{\downarrow
}_{n}$, $F^{\uparrow}_{n}$ enveloping $F(X)$, with direct implications
on the efficiency of
the algorithm, as we explain analytically
below.


\subsubsection*{$F_a$-example: Only refinement} Here, we need
information only on the marginal variable $\sup X_t$ (and not the whole
of the
continuous path on $[0,T]$) to develop an alternating series for
$F_a(X)$. Thus, it suffices to apply a reduced version of the complete
$\varepsilon$-strong algorithm in Table~\ref{tabstrong} where we only
repeatedly refine the initial intersection layer $\mathcal{I}_{0,T}$ (in
particular, we only refine the layer for the maximum) as described in
Section~\ref{secrefine}
(and never bisect it) to construct $F^{\downarrow}_{n}$, $F^{\uparrow
}_{n}$. In particular,
having defined:
\[
\phi(x) = \mathrm{e}^{-rT}( \mathrm{e}^{\sigma x} - K_S )^{+}
\]
knowing that after $n$ refinements the allowed range for $\sup X_t$ is
$[ U^{\downarrow}_{n},U^{\uparrow}_{n}]$ (with initial
position $[U^{\downarrow}_{0},U^{\uparrow}_{0}] = [ X_0\vee X_T,U]$)
we set
$F^{\downarrow}_{n}= \phi(U^{\downarrow}_{n})$, $F^{\uparrow}_{n}
=\phi(U^{\uparrow}_{n})$.


\subsubsection*{$F_b$-example: Refinement and bisection}

The complete machinery of the $\varepsilon$-strong algorithm in Table
\ref
{tabstrong} is required here as we need to bound a path integral.
Recall that the $n$th step of the algorithm provides the piecewise
constant paths
$X^{\downarrow}_{t}(n)$, $X^{\uparrow}_{t}(n)$ enveloping $X$ defined
in (\ref{equplo}). We
now set
$F^{\downarrow}_{n} = F_b(X^{\downarrow}(n))$, $F^{\uparrow}_{n}
= F_b(X^{\uparrow}(n))$.

\subsubsection*{$F_c$-example: Selective refinement and bisection}
We will now only need to bisect a \textit{selection} of intersection
layers as we will be allowed to delete intersection layers that cannot
definitely contain $\sup(\frac{r}{\sigma} t + X_t)$ during the
execution of the $\varepsilon$-strong algorithm. In particular, assuming
the current collection (after $n-1$ steps) of stored intersection layers
$\mathcal{I}_{s_{i},t_{i}}$, with $s_i < t_i \le s_{i+1}$, containing
information about the path $X$, and determining the allowed range for
$\sup(\frac{r}{\sigma} t + X_t)$ :
%
%
\begin{equation}
\label{eqselect}
[ U^{\downarrow}_{n-1},U^{\uparrow}_{n-1} ] = \biggl[ \sup_{i}\biggl\{
U^{\downarrow}_{s_i,t_i} + \frac
{r}{\sigma} s_i\biggr\},
\sup_{i}\biggl\{U^{\uparrow}_{s_i,t_i}+\frac{r}{\sigma} t_i\biggr\} \biggr]
\end{equation}
we proceed to the $n$th step where: (i) we bisect and refine
all stored intersection layers $\mathcal{I}_{s_{i},t_{i}}$, (ii)~calculate
the running bounds $[ U^{\downarrow}_{n},U^{\uparrow}_{n} ]$
by taking the suprema as in (\ref{eqselect}) but now over all newly
obtained intersection layers, (iii) \textit{delete}
the obtained intersection layers $\mathcal{I}_{s,t}$ for which
$U^{\uparrow}_{s,t}<U^{\downarrow}_{n}$ (as they cannot offer extra
information on the whereabouts of
$\sup(\frac{r}{\sigma} t + X_t)$ given that we already know that the
latter is within $[ U^{\downarrow}_{n},U^{\uparrow}_{n} ]$) and store
\textit{only} the
remaining ones for the next iteration.
At each step, we set $F^{\downarrow}_{n}= \phi(U^{\downarrow}_{n})$,
$F^{\uparrow}_{n} =\phi(U^{\uparrow}_{n})$
with $\phi$ as defined above.

%

\subsubsection*{Numerics}
We have run the $\varepsilon$-strong algorithm for the above scenaria. To
give an idea of its computing cost,
we compare its execution times with those of the standard (Euler)
approximation method that replaces the
continuous-time path $\{S_t ; t\in[0,T]\}$ with its discretised
approximation $\{S_{t_i}\}_{i=0}^{l}$, for
$t_i = t_{i-1}+\delta$, with step-size $\delta= T/l$; then,
continuous-time maxima and integrals appearing in
the functionals (\ref{eqF1})--(\ref{eqF3}) are replaced with their
obvious approximations based on the
discrete-time vector $\{S_{t_i}\}$. We run our simulations under the
parameter selections:
\[
r=0.05 ,\qquad \sigma=0.2 , \qquad S_0=1 ,\qquad K=1 ,\qquad T=1 ,\qquad U=1.25 ,\qquad L=0.75 .
\]
Tables~\ref{tabF1}--\ref{tabF3} show results from the simulation
study. For each different algorithm,
we show its execution time (all algorithms were coded in $\texttt
{C++}$) and a 95\% confidence interval for the mean of the realised
estimators to give
an idea about the variance of the estimates and their bias (for the
case of the Euler approximation, as the
$\varepsilon$-strong algorithm is unbiased). The results in Tables~\ref
{tabF1},~\ref{tabF3} are obtained via 100\,000
independent realizations of the estimators, whereas those in Table \ref
{tabF2} via 10\,000 independent realizations.
%

%
\begin{table}
\caption{Simulation results from the application of the Euler
approximation and the $\varepsilon$-strong
algorithm for the estimation of the option price in $\mathrm{E}[
F_a(S) ]$
in (\protect\ref{eqF1}). The results in the table correspond to a sample of
$100\,000$ estimates. $\delta$ is the discretisation increment of the
Euler method,
and $n_0$ is the number of preliminary steps for the $\varepsilon$-strong
algorithm before the simulation of the uniform random variable (see
(\protect\ref{eqimproved}))}\label{tabF1}
\begin{tabular*}{\textwidth}{@{\extracolsep{\fill}}cc@{}}
\begin{tabular*}{187pt}{@{\extracolsep{\fill}}lll@{}}
\multicolumn{3}{@{}l}{Euler approximation}\\
\hline
$\delta$ & Time (secs) & 95\% Conf. Int. \\
\hline
1$/$10 & 0.4 & $[638,647]\times10^{-4}$ \\
1$/$20 & 0.8 & $[657,667]\times10^{-4}$\\
1$/$40 & 1.5 & $[669,679]\times10^{-4}$\\
1$/$80 & 2.9 & $[674,683]\times10^{-4}$\\
1$/$160 & 5.8 & $[680,689]\times10^{-4}$ \\
\hline
\end{tabular*}
&
\begin{tabular*}{180pt}{@{\extracolsep{\fill}}lll@{}}
\multicolumn{3}{@{}l}{$\varepsilon$-strong}\\
\hline
$n_0$ & Time (secs) & 95\% Conf. Int. \\
\hline
2 & 1.1 & $[683,693]\times10^{-4}$ \\
\hline
\end{tabular*}
\end{tabular*}
\end{table}

%
%
\begin{table}
\caption{Similar results as for Table \protect\ref{tabF1}, but now for the
case of the Asian option
$\mathrm{E}[ F_b(S) ]$ in~(\protect\ref{eqF2}) -- with the difference that here
the results correspond to a sample of
$10\,000$ estimates}
\label{tabF2}
\begin{tabular*}{\textwidth}{@{\extracolsep{\fill}}cc@{}}
\begin{tabular*}{187pt}{@{\extracolsep{\fill}}lll@{}}
\multicolumn{3}{@{}l}{Euler approximation}\\
\hline
$\delta$ & Time (secs) & 95\% Conf. Int. \\
\hline
$10^{-1}$ & \phantom{00}0.1 & $[157,169]\times10^{-4}$ \\
$10^{-2}$ & \phantom{00}0.4 & $[120,130]\times10^{-4}$\\
$10^{-3}$ & \phantom{00}3.5 & $[106,116]\times10^{-4}$\\
$10^{-4}$ &\phantom{0}34.3 & $[107,116]\times10^{-4}$\\
$10^{-5}$ & 344.8 & $[102,112]\times10^{-4}$ \\
\hline
\end{tabular*}
&
\begin{tabular*}{180pt}{@{\extracolsep{\fill}}lll@{}}
\multicolumn{3}{@{}l}{$\varepsilon$-strong}\\
\hline
$n_0$ & Time (secs) & 95\% Conf. Int. \\
\hline
2 & 115.9 & $[81,128]\times10^{-4}$ \\
\hline
\end{tabular*}
\end{tabular*}
\end{table}

%
%
\begin{table}[b]
\caption{Similar results as for Table \protect\ref{tabF1}, but now for the
case of
$\mathrm{E}[ F_c(S) ]$ in (\protect\ref{eqF2})}
\label{tabF3}
\begin{tabular*}{\textwidth}{@{\extracolsep{\fill}}cc@{}}
\begin{tabular*}{187pt}{@{\extracolsep{\fill}}lll@{}}
\multicolumn{3}{@{}l}{Euler approximation}\\
\hline
$\delta$ & Time (secs) & 95\% Conf. Int. \\
\hline

$1/10$ & 0.5 & $[797,807] \times10^{-4}$ \\
$1/20$ & 0.9 & $[822,832]\times10^{-4}$\\
$1/40$ & 1.7 & $[833,844]\times10^{-4}$\\
$1/80$ & 3.3 & $[835,846]\times10^{-4}$\\
$1/160$ & 6.6 & $[846,858]\times10^{-4}$ \\
\hline
\end{tabular*}
&
\begin{tabular*}{180pt}{@{\extracolsep{\fill}}lll@{}}
\multicolumn{3}{@{}l}{$\varepsilon$-strong}\\
\hline
$n_0$ & Time (secs) & 95\% Conf. Int. \\
\hline
2 & 178.0 & $[842,854]\times10^{-4}$ \\
\hline
\end{tabular*}
\end{tabular*}
\end{table}

Looking at the three tables, we can make some comments; we focus more
here on giving a simple picture to the reader than
being mathematically precise. The cost \textit{per
sample} of the $\varepsilon$-strong algorithm
corresponds to that of the Euler approximation
with $\delta\approx1/40$, $\delta\approx4^{-1}\cdot10^{-4}$ and
$\delta\approx10^{-3}$ for the cases of
$\mathrm{E}[ F_a(S) ]$, $\mathrm{E}[ F_b(S) ]$ and $\mathrm{E}[
F_c(S) ]$ respectively.
Taking also the standard deviation under consideration (but \textit{not}
the bias)
from the column with the confidence intervals,
for the case of $\mathrm{E}[ F_b(S) ]$ we would need about 25 times more
samples than then Euler approximation
to attain the same range for the confidence interval; thus, ignoring
the bias for the Euler approach, one could say that the overall
cost of the $\varepsilon$-strong algorithm for the case of $\mathrm{E}[
F_b(S) ]$
corresponds to that of the Euler method with step-size
$\delta' \approx(4\cdot25)^{-1}10^{-4}=10^{-6}$.

However, a general remark here is that the $\varepsilon$-strong algorithm
returns unbiased estimators
of the relevant path expectations, and for the applications we have
considered above it can provide
accurate, unbiased estimates in reasonable amounts of time. Even when
ignoring the bias of the Euler approach,
for the cases of $\mathrm{E}[ F_a(S) ]$ and $\mathrm{E}[ F_c(S) ]$
the cost
of the $\varepsilon$-strong algorithm
already seems to be on a par with that of the Euler method for
relatively non-conservative choices of discretisation step $\delta$.
(We should also stress that there is definitely great space for
improving the efficiency of the
used computing code for the $\varepsilon$-strong algorithm.)

\subsection{\texorpdfstring{Remark on number of bisections for $\varepsilon$-strong algorithm}
{Remark on number of bisections for epsilon-strong algorithm}}
\label{secinf}

We make a comment here on the number of required iterations before the
value of the binary variable
$\mathrm{I}_{ F(X)>R}$ in (\ref{eqimproved}) is decided. Proposition
\ref{prl1} will be of relevance in
this context. Recall that $\kappa$ in (\ref{eqkappa}) denotes the
number of steps to decide about $\mathrm{I}_{F(X)>R}$. The cost of
$\kappa$ iterations of the $\varepsilon$-strong algorithm (when its full
machinery
is required) is proportional to
$\mathcal{K} = 2^{\kappa}$.
In the context of (\ref{eqimproved}), we find:
\[
\mathrm{P} [ \mathcal{K} > 2^{n} ] = \mathrm{E}\bigl[ \mathrm{P} [
\kappa
> n\vert X ] \bigr] = \mathrm{E}\biggl[ \frac{F^{\uparrow
}_{n}-F^{\downarrow}_{n}}{F^{\uparrow}_{n_0}-F^{\downarrow}_{n_0}}
\biggr] .
\]
Proposition~\ref{prl1} states that $|X^{\uparrow}(n)-X^{\downarrow
}(n)|_{L_1}=\mathcal
{O}(2^{-n/2})$.
The same rate of convergence will many times also be true for $\mathrm{E}[
F^{\uparrow}_{n}-F^{\downarrow}_{n} ]$:
this will be the case for instance when $F(X)=f(\int_{0}^{1}g(X_s)\,\mathrm{d}s)$
under general
assumptions on $f,g$ (e.g., if $|f(y)-f(x)|\le M(x,y)|y-x|$, for a polynomial
$M$, and the same for $g$; a proof is not essential here).
For such a rate (and since the user-specified $F^{\uparrow
}_{n_0}-F^{\downarrow}_{n_0}$
should be easily controlled),
we will get:
\[
\mathrm{P} [ \mathcal{K} > 2^{n} ] = \mathcal{O}(2^{-n/2})
\]
giving the infinite expectation $\mathrm{E}[ \mathcal{K} ]=\infty$.

In a given application though, one could fix a big enough maximum number
$n_{\mathrm{max}}$, stop the bisections if that number has been reached and
report, say,
$(F^{\uparrow}_{n_{0}}+F^{\downarrow}_{n_0})/2$ as the realization of
the estimator if that happens,
without practical effect on the results. To explain this,
note that we know, from (\ref{eqimproved}),
that the actual unbiased value is either $F^{\uparrow}_{n_{0}}$ or
$F^{\downarrow}_{n_0}$,
so we know \textit{precisely}
that the absolute bias from the single realization when $n_{\mathrm{max}}$ was
reached cannot be greater than $(F^{\uparrow}_{n_{0}}-F^{\downarrow
}_{n_0})/2$.
In total, when averaging over a number of realizations we can have a
precise arithmetic bound on the absolute
value of the bias of the reported average; if $n_{\max}$ is `big
enough' so that we reach $n_{\max}$ only in a small
proportion of realizations the (analytically known) bias could be of
such a magnitude that the reported results
will be precisely the same as when implementing the regular algorithm
without $n_{\mathrm{max}}$ for a reasonably selected degree of accuracy. For
example, in the case of the estimation of $\mathrm{E}[ F_c(S) ]$ in Table
\ref{tabF2},
we have in fact used $n_{\mathrm{max}}=10$ and found that the introduced bias
was smaller than $3\times10^{-5}$ so avoiding
it would not make any difference or whatsoever at the results reported
right now in Table~\ref{tabF2}.

Note that such an issue did not arise in the cases of $\mathrm{E}[ F_a(S)
]$ and $\mathrm{E}[ F_c(S) ]$ when a reduced
version of the $\varepsilon$-strong algorithm was applied.


\section{Further directions for applications}
\label{secexother}

We sketch here some other potential applications of the $\varepsilon
$-strong algorithm.

In the case of barrier options, sometimes
one needs to evaluate expectations involving a Brownian hitting time
(see, e.g.,~\cite{robe97}).
Given a nonconstant boundary $H\dvtx [0,\infty)\rightarrow\mathbb{R}$,
such that
$S_0<H_0$, consider:
\[
\tau_{_H} = \inf\{t\ge0\dvt  S_t \ge H_t \} ,
\]
with $S=\{S_t\}$ being the geometric BM in (\ref{eqgeom}).
The price of a related derivative will be $\mathrm{E}[ F(S) ]$ where now:
\[
F(S) = \psi(S_T)\cdot\mathrm{I}_{ \tau_{_H}<T}
\]
for some pay-off function $\psi(\cdot)$.
This estimator requires the evaluation of $\mathrm{I}_{ \tau_{_H}<T}$
for a realised path.
Such an evaluation is possible under our simulation methods,
since for a given bridge, say from $S_s$ to $S_t$ with $s<t<T$, we can decide
if its maximum is within $[H^{\downarrow}_{s,t},H^{\uparrow}_{s,t}]$ or
not (thus, deciding also whether there is a chance
that the bridge hits $H$ on $[s,t]$ or not), with
\[
H^{\downarrow}_{s,t}=\inf\{H_u;u\in[s,t]\},\qquad H^{\uparrow}_{s,t}=\sup
\{H_u;u\in[s,t]\} ,
\]
using the refinement procedure described in Section~\ref{secrefine}
(more particularly,
a slightly modified version of it, where instead of halving the allowed
variation for the maximum,
it decides if it lies in a given interval or not). Computational effort
will then \textit{only} be spent
on the bridges for which the maximum is indeed in $[H^{\downarrow
}_{s,t},H^{\uparrow}_{s,t}]$,
iteratively bisecting them until a definite decision is reached about
whether $H$ has been hit.

Individual simulation techniques employed in the development
of the $\varepsilon$-strong algorithm are also of independent interest.
For instance, we have exploited during the construction of the
$\varepsilon
$-strong algorithm
a monotonic property at the core of
the Brownian structure; we can further use this characteristic to
develop original simulation
techniques for Brownian distributions.
One application for instance could involve dynamics of Brownian motion
restricted to stay in a bounded
domain. A Brownian motion with constant drift, restricted to remain in
$(-\uppi/2,\uppi/2)$,
is known (see~\cite{pins85}) to be described via the stochastic
differential equation:
%
%
\begin{equation}
\label{eqtan}
\mathrm{d}X_t = - \tan(X_t)\,\mathrm{d}t + \mathrm{d}W_t .
\end{equation}
Unbiased sampling methods for $X_t$ are not (to the best of our
knowledge) available;
one has to resort to Euler, or other, approximations.
We can, however, now construct an exact sampling algorithm based on the
methods so far described.
Girsanov's theorem provides the following expression for
the transition density of the Markov process (\ref{eqtan}):
\begin{eqnarray}\label{eqtanx}
p(y;x,t)&:=&\mathrm{P} [ X_t\in \mathrm{d}y\vert X_0=x ] / \mathrm{d}y
\nonumber
\\[-8pt]
\\[-8pt]
\nonumber
&\phantom{:}=& \frac{\cos(y)}{\cos(x)}
\gamma(-\uppi/2,\uppi/2;t,x,y) p_0(y;x,t),\qquad y\in(-\uppi/2,\uppi/2) ,
\end{eqnarray}
for the unconditional Brownian transition density:
\[
p_0(y;x,t) = (2\uppi t)^{-1/2}\mathrm{e}^{-(y-x)^2/(2t)} .
\]
Density (\ref{eqtanx}) has a structure reminiscent of that of the
density of the middle point in Proposition~\ref{prmiddle}:
ideas employed there, are also relevant now. Analytically,
for $x$ not close to the boundaries, one can simply carry out a
rejection sampler
with proposals from $p_0$. Then, the acceptance/rejection decision will
be based on
comparing a real number with $\gamma(-\uppi/2,\uppi/2;t,x,y)$ following the
pattern described in Section~\ref{seczeta}.
As $x$ approaches the boundaries, this algorithm becomes inefficient.
But, similarly to the method for the simulation of the density in
Proposition~\ref{prmiddle},
partial sums from the infinite series-expression for $\gamma(-\uppi
/2,\uppi
/2;t,x,y)$
can be incorporated in the proposal to produce an efficient algorithm
(Section~\ref{subsecmiddle}
describes the algorithm for the more complex density appearing there;
here, we omit the details).

%

\section{Conclusions}
\label{secconclusion}
We have presented a contribution to sampling methods for Brownian dynamics:
a new iterative algorithm that envelopes the Brownian path, thus offering
explicit information for all its aspects (minimum, maximum, hitting times).
Individual steps of the algorithm could be of independent interest,
yielding new
sampling methods for distributions derived from Brownian motion dynamics.

We should remark here on the generality of our scope.
The $\varepsilon$-strong algorithm (or some of its individual steps)
can provide, more or less unchanged,
unbiased Monte Carlo estimators
in separate estimation problems, for which quite an extensive amount
of case-specific methods have been investigated in the literature;
for instance,
one can refer to the long literature for
the applications we briefly described in Sections~\ref{secexpaths}
and~\ref{secexother}.

We have presented some applications and sketched some others towards
illustrating the potential of our methods.
Note that the infinite expectation issue remarked in Section~\ref{secinf}
is a direct consequence of the Brownian dynamics: the maximum
of the Brownian path scales as $\Delta t^{1/2}$ on a small time
interval $[0,\Delta t]$
(see, e.g.,~\cite{revu99}). Thus, any enfolding processes will
necessarily converge not
faster than $\mathcal{O}(\Delta t^{1/2})$ (which is the order attained
by the $\varepsilon$-strong algorithm).
This relatively slow convergence of the enfolding processes also
explains the increased cost for when
estimating $\mathrm{E}[ F_b(S) ]$ in Section~\ref{subsecnum}.
We envisage that it might be possible to combine the iterative process
of the algorithm with
a \textit{coupling} step once the bounding processes are relatively close
to each other to
overcome this long anticipation (in the spirit of~\cite{robe02,besk05}).
We hope to formalise this idea in future research.

\section*{Acknowledgements}
We thank the referees for many valuable comments and suggestions that
have greatly improved the content of the paper.

%
%
%

\printhistory

\end{document}